\documentclass[prd,aps,floats,twocolumn,nofootinbib]{revtex4-1}
\usepackage{slashed}
\usepackage{mathtools}
\usepackage{amsfonts}
\usepackage{amssymb}
\usepackage{epsfig}
\usepackage{empheq}
\usepackage{mathrsfs}
\usepackage{xcolor}

\usepackage{bm}

\usepackage{appendix}

\begin{document}
\newcommand*\widefbox[1]{\fbox{\hspace{2em}#1\hspace{2em}}}
\newcommand{\m}[1]{\mathcal{#1}}
\newcommand{\nn}{\nonumber}
\newcommand{\ph}{\phantom}
\newcommand{\eps}{\epsilon}
\newcommand{\be}{\begin{equation}}
\newcommand{\ee}{\end{equation}}
\newcommand{\bea}{\begin{eqnarray}}
\newcommand{\eea}{\end{eqnarray}}
\newcommand{\cH}{{\cal H}}
\newtheorem{conj}{Conjecture}

\newcommand{\plk}{\mathfrak{h}}


\title{
Lorentz Violation in Emergent Gravity and Its Cosmological Consequences}

\date{}

\author{Raymond Isichei}
\author{Jo\~{a}o Magueijo}
\email{magueijo@ic.ac.uk}

\affiliation{Abdus Salam Centre for Theoretical Physics, Imperial College London, Prince Consort Rd., London, SW7 2BZ, United Kingdom}

\begin{abstract}
We show that General Relativity and other geometrical theories can be viewed as a degenerate Otto cycle with only heat-exchange legs in emergent gravity. Including work-producing legs yields controlled violations of local Lorentz invariance and energy–momentum conservation, which  produce late-time cosmological acceleration. Implications for the cosmological constant problem, structure formation and local observations are discussed.
\end{abstract}

\date{26 Nov 2025}

\maketitle


The view that gravity may be an emergent phenomenon arising from the thermodynamic limit of microscopic space-time structures dates back to~\cite{J} and has since been extensively explored~\cite{Jnoneq,ChircoLiberati2010,Paddy1,Paddy2,Paddy3,Paddy4,V1,V2,J2}. 
By inverting Bekenstein’s~\cite{Bek} and Hawking’s~\cite{Hawk} result that space-times endowed with horizons possess entropy $S$ and temperature $T$, one finds that the relation for the internal energy $\delta U = T \delta S$, universally applied to a local Rindler horizon, implies the Einstein equations (or rather, these appear as an equation of state). We refer to this framework as {\it thermogravity}. The question is whether thermogravity can also generate genuinely new theories, beyond General Relativity (GR) and its known extensions~\cite{Paddy1},  whose macroscopic signatures directly reflect microscopic structure. Such a possibility is important both as a model-independent bridge between observation and quantum gravity and as a response to concerns that thermogravity is a tautology or an accident. If it yields distinctive theories with testable phenomenology, its case becomes substantially stronger.

We investigate this matter along a simple line.
Setting aside secondary complications (such as the volume terms in~\cite{Paddy1,Cliff1,Cliff2,VisserEngine,Hayward,CaiCao}),
the striking feature of thermogravity is that  GR and its geometrical extensions arise purely from  heat and its pair of variables, $T$ and $S$. 
Their bare bones require no additional thermodynamic pairs of variables with an associated ``work'' and Carnot-like cycles. This observation alone 
provides a hint: if such pairs of variables were to exist, one variable in each pair would have to be held constant in standard theories, so that these theories appear as a section of the larger system. 
What could these variables be, and 
what new theories might follow from allowing them to vary? 

There are several possibilities. One is that 
they appear as effective parameters in the emergent macroscopic description, in which case their required constancy in standard theory identifies half of them with the fundamental constants of Nature~\cite{thermoconsts}. In this Letter we follow an alternative route where they are coarse-grained averages of inaccessible microscopic quantum-gravitational structures, for instance, the density of sprinkling events in causal set theory~\cite{BombelliLeeMeyerSorkin1987,Sorkin2003,Surya2019}, spin-network nodes in loop quantum gravity~\cite{AshtekarLewandowski2004,Thiemann2007,
RovelliVidotto2014,Perez2013}, or “atoms of space” in deformed special relativity frameworks~\cite{AmelinoCamelia2002,MagueijoSmolin2002,KowalskiGlikman2005}. If such microstructures carry a conserved ``number'' $N$, they introduce a ``chemical potential'' $\mu$, contributing a new $\mu dN$ term to $\delta U$. We show that the new “work” terms can induce controlled violations of Lorentz invariance, reflecting the preferred frame inherent in most thermodynamic settings. These lead to corresponding violations of stress–energy conservation, with implications for the cosmological constant problem and the observed cosmic acceleration—both potentially imprints of the emergent nature of local Lorentz symmetry.

Before embarking on our derivation, we stress that thermogravity departs radically from the usual construction of a theory via an action, equations of motion, symmetries,  conserved quantities, etc. The fundamental action is hidden in the microstructure, and there is no {\it a priori} reason why the emergent dynamics should take the form of PDEs decoupled from thermodynamics, much less admit an action and gauge symmetries. 
Thus, we have the latitude to explore theories that would be anathema from the traditional point of view.
Moreover, in this context, thermodynamic energy is more fundamental than the external notion of energy in curved spacetime, with all its ambiguities of definition (a point that will be relevant when discussing local symmetries and Killing fields). Finally, the internal thermodynamical time could bear no relation to the outer time of the emergent spacetime~\cite{FN+}.

\begin{figure}
  \centering
  \includegraphics[width=0.2
  \textwidth]{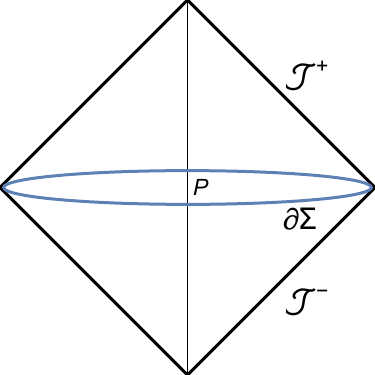}
  \caption{Point $P$ (and a section of its worldline) and light cones $\mathscr{I}^\pm$ and mirror $\partial\Sigma$ defining the causal diamond.}
  \label{FigDiamond}
\end{figure}

Other settings could be used to present our argument, but for definiteness we will consider a small causal diamond $\mathcal{D}$ around a generic point $P$ (see Fig.\ref{FigDiamond}): a neighborhood of $P$ bounded by a lower ($\mathscr{I}^-$) and an upper ($\mathscr{I}^+$) null surface of infinitesimal affine length $2 \ell$, joined at an “equator’’ (mathematically, a bifurcation surface) which is the boundary $\partial\Sigma$ of a three-ball $\Sigma$. Physically, this is the region traced by an expanding sphere of light ($\mathscr{I}^-$), reflecting at $\partial\Sigma$, and then ($\mathscr{I}^+$) reconverging to a point. 
The causal-diamond construction is particularly useful in emergent-gravity frameworks that preserve diffeomorphism and local Lorentz invariance, but it also provides a natural probe for scenarios in which these symmetries are ultimately broken. Notably, the equator breaks local Lorentz invariance, and one can consider infinitely many such spheres, boosted with respect to one another, giving rise to physically distinct diamonds. In contrast to~\cite{J2}, the mirror here is attached to a preferred frame, with local Lorentz invariance emerging only if the mirror has no physical macroscopic effects~\cite{FN+++}.

\begin{figure}
  \centering
  \includegraphics[width=0.5\textwidth]{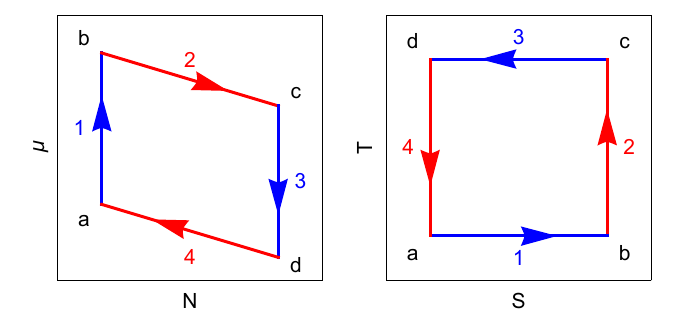}
  \caption{Sketch of an infinitesimal Otto cycle in the $(N,\mu)$ and $(S,T)$ planes. GR and its geometric relatives arise from a degenerate cycle, obtained by collapsing the red legs to a point and retaining the  blue legs (corresponding to the fluxes through $\mathscr{I}^{\mp}$). Adding the isentropic red legs leads to the proposed new theories, for which the heat in and out are not the same, and ``chemical'' work is done.}
  \label{FigOtto}
\end{figure}

For the purposes of this paper, the essential feature of the diamond is that energy can only flow inward across $\mathscr{I}^-$ and outward across $\mathscr{I}^+$, and that all the entropy is on $\mathscr{I}^\pm$~\cite{FN1},  so that such flows are heat flows (this could be used as the emergent definition of ``null surface''). The idea is to subject $\mathcal{D}$ to an Otto cycle such that GR and relatives appear as the iso-$N$ (the ``isochoric'') no-work limit, and the sought-after new theories emerge from the generic cycle (Fig.~\ref{FigOtto}). 
A form of Stokes' theorem should then enforce:
\begin{equation}\label{stokes}
    0=\delta U_{tot}=\int _{\mathscr{I}^-}\delta U-\int _{\mathscr{I}^+}\delta U+\int _{\mathcal{D}}\delta U=\Delta Q +W
\end{equation}
with the boundary fluxes (the heat $Q$) counted as inward/outward on $\mathscr{I}^\mp$, and the last term (the ``chemical work'', $W$) unwrapped into two work contributions that return the system to its initial state. In this way, the diamond naturally encodes a cycle.

The advantage of a diamond is that it creates a compact region bounded by a null surface (except for the bifurcation, which has measure zero), so that Stokes' theorem can be applied. The price to pay is that $\mathcal{D}$ is invariant under only a {\it conformal} Killing vector (KV) field~\cite{J2}:
\begin{equation}\label{CKVF}
    \zeta  =\frac{\kappa}{2\ell}[(\ell ^2-r^2-t^2)\partial_t -2rt\partial_r]
\end{equation}
where $\pounds_\zeta    \eta_{\mu\nu}\equiv \nabla_\mu \zeta_\nu+\nabla_\nu \zeta_\mu=2\psi\eta_{\mu\nu}=-\frac{2\kappa t}{\ell}\eta_{\mu\nu}$ and  $\kappa$ is the surface gravity. This is merely a technical problem with $\mathcal{D}$: the fact that the internal thermodynamic energy is more fundamental implies that the emergent external theory must have a local Killing vector $\chi$. In general $\chi$ and $\zeta$ are not simply related, but at the diamond center (or more generally at $t=0$) and on $\mathscr{I}^\pm$ they are proportional. Setting 
$\bar u=u+\ell=t-r+\ell $ and $\bar v=v-\ell=t+r-\ell$ we have:
\begin{equation}
    \zeta^\mu=\frac{\kappa}{2\ell}[(2\bar u\ell-\bar u^2)\partial_{\bar u}+(-2 \bar v\ell-\bar v ^2)\partial_{\bar v}]
\end{equation}
with $\bar u=0$ and $-2\ell\le \bar v\le 0 $ on $\mathscr{I}^-$, and $\bar v=0$ and $0\le \bar u\le 2\ell$ on $\mathscr{I}^+$. Thus, $\bar u$ and $\bar v$ are affine parameters $\lambda$, set to zero at the bifurcation and increasing to the future (as in the conventions in~\cite{J}), so that $\chi=\pm \kappa \lambda \partial_{\lambda}$ on $\mathscr{I}^\pm$. We then have $ \zeta^\mu=\left(1\mp\frac{\lambda}{2\ell}\right)\chi^\mu$ on $\mathscr{I^\pm}$.

The ``isochoric'' assumption on legs 1 and 3 (i.e. heat only, on $\mathscr{I}^\mp$) ensures a reduced set of Einstein equations. This can be obtained (see~\cite{J} for more detail) by writing the energy flux (adjusting signs as appropriate for $\mathscr{I}^\pm$) as $\delta U=P^\mu d\Sigma_\mu$, where $P^\mu$ is the $4$-momentum (defined from the stress-energy tensor $T_{\mu\nu}$ and the local Killing vector as $P^\mu=T^\mu_{\;\;\nu}\chi^\nu$) and $d\Sigma_\mu$ is the area element vector. We then write $d\Sigma_\mu=k^\mu d\lambda\,dA$, where $dA$ is the spatial cross section of the element and $k^\mu$ is a generator. The variation in the entropy for element $dA$ is $\delta S=\delta dA/L_1^2$ (with $L_1$ a length scale), and the Raychaudhuri equation~\cite{J} for the expansion of geodesics gives to leading order $\delta dA\approx \lambda R_{\mu\nu}k^\mu k^\nu d\lambda dA$ (where $R_{\mu\nu}$ is the Ricci tensor). Imposing $\delta U = T \delta S$ on {\it each} surface element of $\mathscr{I}$ finally yields:
\begin{equation}\label{Eeqn}
\left(\kappa T_{\mu\nu}-\tfrac{T}{L_1^2}G_{\mu\nu}\right)k^\mu k^\nu=0
\end{equation}
for any null $k^\mu$ (where the Einstein tensor is $G_{\mu\nu}=R_{\mu\nu}-\frac{1}{2}g_{\mu\nu}R$, and so $R_{\mu\nu}k^\mu k^\nu=G_{\mu\nu}k^\mu k^\nu$)~\cite{FN2}. 

These are the Einstein equations, subject to a few qualifications. 
First, we must require that $L_1$ be independent of $\kappa$, so that the gravitational constant $G_N$ is independent of the temperature (with $T=\kappa\hbar /2\pi$ and $L_1=2\sqrt{\hbar G_N}$). Only then do the field equations decouple from the underlying thermodynamics; but this
implies that $S$ is independent of $T$, an odd but common assumption, which we will keep for the time being. Second, what we do get is 
the {\it traceless} Einstein equations,
\begin{equation}\label{Eqenstracefree}
G_{\mu\nu}-\tfrac{1}{4}g_{\mu\nu}G=8\pi G_N \bigl(T_{\mu\nu}-\tfrac{1}{4}g_{\mu\nu}T\bigr),
\end{equation}
i.e. unimodular gravity. This integrates to Einstein’s equations with a cosmological constant, but only under 
the additional assumption that $\nabla_\mu T^{\mu\nu}=0$, 
which we are about to abandon.

Finally, we could consider implementations arising from applying $\delta U=T\delta S$ to structures on $\partial \mathcal{D}$ larger than each surface element.  In the most extreme case, we might impose it only on the whole $\mathscr{I}^\pm$, leading to ``light-cone averaged'' Einstein equations
\begin{equation}\label{lightconeE}
\langle k^\mu k^\nu G_{\mu\nu }\rangle_{l.c.}=8\pi G_N \langle k^\mu k^\nu T_{\mu\nu}\rangle_{l.c.}
\end{equation}
where here and later we adopt the notation
 $   \langle k^\mu k^\nu X_{\mu\nu} \rangle_{l.c.}=\frac{1}{4\pi}\int d^2 \hat{\mathbf{k}}\,  k^\mu k^\nu X_{\mu\nu}$
(with $k^\mu=(1,\hat{\mathbf{k}})$). If 
$n^\mu$ is the mirror's normal, then \cite{SMB}:
\begin{equation}\label{lcavdef}
    \langle k^\mu k^\nu X_{\mu\nu} \rangle_{l.c.}=X_{\mu\nu}n^\mu n^\nu +\frac{1}{3}h^{\mu\nu}X_{\mu\nu}
\end{equation}
where $h^{\mu\nu}=g^{\mu\nu}+n^\mu n^\nu$. Hence, the strongest implementation reduces the ten Einstein equations to nine (unimodular gravity), while the weakest reduces them to a single light-cone averaged equation (breaking local Lorentz invariance). The more Einstein equations we lose, the more additional thermodynamic equations of state are required, up to a maximum of nine. 
This will be essential to what follows. 

We now consider a more general infinitesimal Otto cycle, 
with isochoric (iso-$N$) legs 1 and 3, and adiabatic (iso-$S$) legs 2 and 4, following from internal energy
\begin{equation}
    U=U(S,N)\implies \delta U=T\delta S +\mu \delta N.
\end{equation}
Secondary equations of state arise from
$T=\left(\tfrac{\partial U}{\partial S}\right)_N$ and $ \mu=\left(\frac{\partial U}{\partial N}\right)_{S}$, constrained by Maxwell's relations,
 $   \left(\frac{\partial T}{\partial N }\right)_{S}=\left(\frac{\partial  \mu }{\partial S}\right)_N$.
For definiteness in Fig.~\ref{FigOtto} we assumed  $\tfrac{\partial T}{\partial N}=\tfrac{\partial \mu}{\partial S}>0$ and a cycle with $W>0$ (but see~\cite{SMA} for other options).
The net heat entering the diamond (which vanishes in standard thermogravity) is carried only by legs 1 and 3,
\begin{align}\label{Q}
\Delta Q&= 
\delta Q_1-\delta Q_3=  \Delta T\delta S\approx - \left(\frac{\partial T}{\partial N}\right)_S\,  \delta N \delta S,
\end{align}
where $\Delta T=T_1-T_3<0$  and $\delta S=\delta S_1 = |\delta S_3|$ (note $\Delta S=\delta S_1-|\delta S_3|=0$). 
The chemical work done in steps 2 and 4 is:
\begin{align}\label{Work}
    W&=
    \delta U_2-\delta U_4=
    \frac{\mu_c+\mu_b}{2}\delta N-
     \frac{\mu_a+\mu_d}{2}\delta N\nn\\
&=    \Delta\mu\delta N \approx \left(\frac{\partial\mu }{\partial S}\right)_N\delta S \delta N,
\end{align}
where $\delta N=\delta N_2 = |\delta N_4|$ (from $\Delta N=\delta N_2-|\delta N_4|=0$) and $\Delta \mu=\delta \mu _1\approx  |\delta \mu_3|$. 
Since $U$ is an exact form, $\Delta Q+W=0$, consistent with Maxwell's relations. For an infinitesimal cycle, $\Delta Q\approx \delta Q_1 \Delta T/T$ (with $T\approx T_1\approx T_2$), so the cycle efficiency satisfies
\begin{equation}
    \eta=\frac{W}{\delta Q_1}=-\frac{\Delta Q}{\delta Q_1}\approx -\frac{\Delta T}{T}
    \approx \frac{\delta N}{T}\frac{\partial T}{\partial N}
    =\frac{\delta N}{T}\frac{\partial \mu}{\partial S}.
\end{equation}
In contrast with the conventional Otto cycle, the efficiency here depends on the shape of the cycle: the degenerate cycle (leading to GR) has zero efficiency, while for $\delta N\neq 0$ and $\partial T/\partial N\neq 0$ (implying $\Delta T\neq 0$) it does not. 
We cannot appeal to Euler or Gibbs-Duhem relations for this system, but instead we will require that $\eta$ scales linearly with $\ell$, so that:
\begin{align}\label{etaeqstat}
    \eta\equiv \frac{W}{\delta Q_1}
    =\frac{2}{15} \frac{\ell}{L_2}\approx -\frac{\Delta T}{T}
\end{align}
with $L_2$ a second length scale, which {\it a priori} can be positive or negative (and a coefficient added for later convenience). This will ensure that the emergent theory decouples from the underlying thermodynamical setup, as we now demonstrate. 

If we take the limit $\ell\rightarrow 0$, then the complications arising from $\mathcal{D}$ being invariant under only a conformal KV disappear. Returning to \eqref{stokes}, we have that indeed Stokes' theorem implies $W+\Delta Q=0$ with~\cite{FN++}:
\begin{align}
    W&=-\int dV [\nabla_\mu(\zeta_\nu T^{\mu\nu}) -\psi T^\mu_\mu]=-\int dV \zeta_\nu\nabla_\mu T^{\mu\nu}\nn\\
    \Delta Q &=-\int _{\mathscr{I}^-}d\Sigma_\mu \zeta_\nu T^{\mu\nu}+\int _{\mathscr{I}^+}d\Sigma_\mu \zeta_\nu T^{\mu\nu}-\int dV \psi T^\mu_\mu\nn.
\end{align}
We subtract from $W$ the correction arising from $\zeta$ being a conformal KV, because (unlike in previous work, e.g.~\cite{Hayward,JacobsonVisser}) we count only as work that which is not part of standard GR and relatives. However, because of the $t\rightarrow -t$ symmetry of the diamond, the correction term is of order $l^5$~\cite{SMB}. Therefore it does not affect: 
\begin{equation}
     \Delta Q=\frac{\Delta T}{T}\delta Q=\frac{\Delta T}{T}\left(\frac{4\pi}{5} \kappa \ell^4  \langle k^\mu k^\nu T_{\mu\nu }\rangle_{l.c.}+\mathcal{O}(\ell^5)\right)\nn
\end{equation}
because we only need $\delta Q\approx \delta Q_1\approx \delta Q_2$ to order $\ell^4$ (see~\cite{SMB} for this and related calculations). 
Evaluating:
\begin{align}
       W&=-\frac{8\pi }{45} \kappa \ell^5  n_\mu \nabla_\nu T^{\mu\nu} +\mathcal{O}(\ell^6)
\end{align}
and combining with \eqref{etaeqstat} (containing the assumed scaling with $\ell$) and \eqref{lcavdef} we finally get:
\begin{align}\label{nonConsTeqn}
      -n_\mu \nabla_\nu T^{\mu\nu}=
      \frac{1}{L_2}\left(T_{\mu\nu}n^\mu n^\nu +\frac{1}{3}h^{\mu\nu}T_{\mu\nu}\right)
\end{align}
to be added to the trace-free Einstein equations \eqref{Eqenstracefree}, to provide the complete set of equations of the theory. 
Since stress--energy conservation is relaxed, there are also controlled violations of diffeomorphism invariance, in addition to local Lorentz invariance~\cite{FNDiffLLIV}.

Hence, the theory involves two equations of state, each tied to a distinct length scale. The first is the familiar Einstein equation of state, characterized by $L_1$. The second introduces a new scale, $L_2$, which governs violations of energy conservation and sets the efficiency of the Otto cycle. While $L_1$ is small (twice the Planck length), $|L_2|$ must be large, ensuring that the new effects remain suppressed in regimes where GR is well tested. It is intriguing to speculate that the scaling $\eta \propto \ell$ (required for decoupling thermodynamics from the emergent theory) may reveal the dimensionality of the microscopic constituents: whether pointlike, string-like, or wall-like.

Let us now apply this theory to 
a homogeneous and isotropic Universe with scale factor $a(t)$ (where $t$ is proper cosmological time),  
spatial curvature $K=0,\pm 1$, density $\rho(t)$ and pressure $p(t)$. The theory's complete equations, \eqref{Eqenstracefree} and \eqref{nonConsTeqn}, simplify to
\begin{align}
    \left(\frac{\dot a}{a}\right)^2+\frac{K}{a^2}-\frac{\ddot a}{a}&= 4\pi G_N(\rho+p),\\
    \dot \rho+3\frac{\dot a}{a}(\rho+p)&=\frac{1}{L_2}(\rho +p),\label{dotrho}
\end{align}
where we have identified the mirror's rest frame with the cosmological frame. General solutions can be derived, and an example (with $p=K=0$) is
\begin{align}
    a&=a_\star t^{2/3} \exp{\left(\frac{t}{3L_2}
    \right)},\label{aoft}\\
    \rho&=\frac{\rho_\star }{a^3}\exp{\left(\frac{t}{L_2}\right)}=\frac{\rho_\star}{a_\star^3}\frac{1}{t^2},\label{rhooft}
\end{align}
where $a_\star$ and $\rho_\star$ are integration constants. It reduces to the usual matter dominated solution for $t\ll |L_2|$, but deviates otherwise, accelerating from $t=(\sqrt 6-2)L_2$ onward, if $L_2$ {\it is positive}. 
Hence, we need a cycle with $W>0$ and $T_3>T_1$, resulting in matter creation (rather than destruction), to generate late-time acceleration. Unfortunately this does not fix the sign of $\tfrac{\partial T}{\partial N}=\tfrac{\partial \mu}{\partial S}$, which remains a microphysical mystery~\cite{SMA}. 

We defer to a future publication a more exhaustive examination of the cosmological phenomenology, stressing that a fully self-consistent assessment is non-trivial. Naively, one expects a distinctive profile for the acceleration as a function of redshift, with $L_2\sim 10^{10}\,{\rm ly}$ required to fit the data. However, a re-examination of the astrophysics involved (for example in building the Hubble diagram) is warranted, because matter creation may be non-negligible in the associated processes. The cosmological fluctuations are also expected to be unique and moreover depend on the choice of $n^\mu$ beyond zeroth order. The preferred frame is physically defined as the frame where matter, but not momentum, is created,
and beyond zeroth order could be anything. Results arising from implementing this theory in the synchronous, comoving and  longitudinal gauges will be different. 
Fluctuations also require further theoretical information pertaining to multi-component systems. 
Indexing components by $I$, Eq.~\eqref{nonConsTeqn} could in general be:
\begin{align}\label{nonConsTeqn1}
      -n_\mu \nabla_\nu T^{\mu\nu}_{I}=\left( n^\mu n^\nu +\frac{1}{3}h^{\mu\nu}\right)
      \sum_JM_{IJ}T_{\mu\nu}^{J} 
\end{align}
with $M_{IJ}=\delta_{IJ}/L_2$ for a minimal model, $M_{IJ}=\delta_{IJ}/L^I_2$ for a generic diagonal model~\cite{FN8}, 
$M_{IJ}=\delta_{I1}/L_2$ for a dark matter model (DM)  where species $J=1$ absorbs all the violations of energy conservation, and others. At its most extreme the DM model could explain the origin of {\it all} the dark matter, with 
  $ \Omega_{DM}/\Omega_b=\exp{\left(t/3L_2\right)}$
(for $K=0$), but this is not needed. Overall, an interesting and uncharted corner of cosmological model building is unveiled.

As with any symmetry-breaking framework, one may restore invariance by introducing Stückelberg fields, but in extreme cases like ours, this is not especially illuminating. As shown in~\cite{SMC}, 
this would map our model into a particularly baroque model of interacting dark energy~\cite{IntDE,Eleanora} subject to Lorentz invariance breaking. By contrast, the direct symmetry-breaking picture offers considerably greater transparency. 
The same applies to the cosmological constant problem, where the symmetry breaking picture opens up new perspectives on both the old~\cite{Weinberg} and new~\cite{Padilla} versions of the problem. It is remarkable that the {\it weakest} requirement needed to accommodate a non-degenerate thermodynamic cycle is to subtract the trace of Einstein’s equations (see \eqref{Eqenstracefree} vs \eqref{lightconeE}). In all implementations, the vacuum energy does not gravitate, so that radiative instabilities are avoided~\cite{Ng,LeeUnimod}. The objection that unimodular-like gravity constitutes nothing more than a gauge-fixed version of GR~\cite{Tonysrant} does not apply to our theory, since it lacks gauge invariance from the outset.

Furthermore, the additional criticism that unimodular-flavored theories merely shift the problem to an integration constant, appears transfigured. In lieu of a finely-tuned integration constant, the theory predicts matter creation and a breakdown of local Lorentz invariance. 
At the level of the macroscopic field theory, this looks like a theory that can never be free. 
A puzzle remains regarding the disparity between the Planck length $L_1$, which sets the gravitational coupling and entropy–area relation, and a new scale $L_2$, which governs energy non-conservation and the efficiency of work production in the Otto cycle. From this viewpoint, the ``new'' cosmological constant problem becomes the smallness of Lorentz-violating effects. Standard thermodynamics naturally prefers a rest frame, and this preference is inherited by thermogravity, but its impact is suppressed by the extreme inefficiency of the Otto cycle we have proposed. Local Lorentz invariance thus emerges as an extraordinarily good  approximate symmetry, and the observed smallness of Lambda-like effects (such as late-time acceleration) becomes a direct reflection of this fact.






So far, we have presented the most minimal theory within this framework, but extensions are natural. Chief among them is the possibility (foreshadowed throughout this Letter) that thermodynamics never decouples from the macroscopic field equations. 
For $T$ and $S$, this decoupling relies on $S$ being independent from  $T$ (or $L_1$ from $\kappa$). Abandoning this assumption makes $G_N$ acquire $T$ dependence, placing the theory squarely within the remit of rainbow gravity~\cite{rainbow}, with the underlying temperature defining the energy scale with which the geometry ``runs''. Legs 1 and 3 would then tilt away from horizontality in the $(S,T)$ plane, but generalizing to the full Otto cycle, our calculation would still go through, with the twist that $L_2$ may itself depend on $\kappa$, producing a simultaneous running $L_i=L_i(T)$. We may then look to the opposite end of the Universe, near the Big Bang and in the UV limit, with
$L_1\sim L_2$ possible. This would imply exponential matter production and acceleration in the very early Universe, with overtones of inflationary and (quasi-)steady state models.

Along another direction we could also have more pairs $\{N_i,\mu_i\}$, additional equations of state and fewer Einstein equations, as suggested in relation to \eqref{lightconeE}. Once homogeneity or isotropy are dropped, this becomes relevant.
Taking the Bianchi I model as an example~\cite{SMD},
the usual four Einstein equations reduce  to three via \eqref{Eqenstracefree}, but by light-cone averaging (see~\eqref{lightconeE}) they further collapse to a single equation. 
The system therefore requires two additional, non-conservation–type equations supplied by the thermodynamics but already in vacuum the solutions differ from the familiar Kasner solutions in GR. Specifically, they can expand in all directions~\cite{SMD}; e.g.,
 $ds^{2} = -dt^{2} + t(dx^{2} + dy^{2}) + t^{2+\sqrt{7}}dz^{2} $.
This reframes the cosmic isotropy problem (unstable in standard theory) as one controlled by the extra equations of state needed to close the system, shifting the burden from the emergent macroscopic theory to the underlying theory.

Finally, in thermodynamics, fixing the ``volume'' (or the \(N_i\)) is not a law of nature but a choice of setup. Likewise, the Otto cycle adopted here or its variations, cannot be elevated to universal laws. Could the thermodynamic setup itself vary across regions of the Universe or across epochs and scales? 
It is perhaps within this broader class of theories, spanning the three lines outlined above, that one should assess the implications for local observations, in tandem with the cosmological ones. With the caveats already emphasized, the latter imply for the most minimal models a baseline \(L_2\sim 10^{10}\,{\rm ly}\). Translating this to the regime of local measurements depends crucially on the local \(n^\mu\) beyond zeroth order (and indeed beyond the linear cosmological--perturbation regime). Furthermore, even with model \eqref{nonConsTeqn1}, for a generic $M_{IJ}$ the $L_{2b}$ for visible (baryonic) matter could be very different from the one inferred from cosmology, $L_{2DM}$.

In the low--speed limit, Lorentz symmetry breaking manifests as a violation of Galilean invariance. Applying Eq.~\eqref{nonConsTeqn} to a point particle \((T_{\mu\nu}=m u_\mu u_\nu,\; n^\mu=(1,\mathbf{0}),\; u^\mu\simeq (1,\mathbf{v}))\) gives at zeroth order \(\dot m/m\simeq c/L_2\), where we reinstate \(c\) for clarity. The spatial components of the conservation equation then yield an anomalous acceleration \(\delta\mathbf{a}=-(\dot m/m)\mathbf{v}\), whose fractional impact is maximized for large velocity-to-acceleration ratios,
\begin{equation}
    \frac{\delta a}{a}\sim \frac{v}{c\,L_2\, a},
\end{equation}
with \(v\) determined by the choice of \(n^\mu\).
Although the conventional Eötvös parameter vanishes, this is an example of equivalence principle violation due to energy non-conservation (see, e.g., Ref.~\cite{EP}), with effects similar to other friction-like phenomena previously explored in the literature~\cite{Orbits}. At {\it face value}, the minimal diagonal $M_{IJ}$ model is ruled out by Solar-System (SS) observations. Unless the preferred frame comoved with all visible matter, planetary orbits would experience a drag; moreover, regardless of \(n^\mu\), the solar mass would increase at a rate \(\sim 10^{-10}\,{\rm yr}^{-1}\). Taken together, these effects would induce orbital changes at levels implying $L_{2b}/L_{2DM}<10^{-3}$.

The non-diagonal ``DM'' model faces similar difficulties, at least superficially.  It implies that all matter moving with respect to the preferred frame, including visible matter, leaves a trail of DM.
If the local $n^\mu$ were to be signaled by the CMB dipole, SS observations would again place similar constraints on the model, since the DM produced by the Sun would remain bound in a disc roughly the size of the solar radius (the CMB-dipole speed is below the escape speed inside the Sun). If, however, the speed relative to the preferred frame were larger, the DM trail could exceed escape velocity and leave negligible imprint on SS dynamics~\cite{FNCerenkov}.

With all the ifs and buts, SS observations appear to require \(L_2/c \gtrsim 10^{13}\,{\rm yr}\) on the relevant scale. By contrast, analogous considerations for galaxy rotation curves (particularly those of the Milky Way) remain compatible with \(L_2/c\sim 10^{10}\,{\rm yr}\). Reconciling cosmological and local constraints thus seems to call for either a running \(L_2\) or  $L_{2b}/L_{2DM}< 10^{-3}$. Given the highlighted uncertainties, we should resist drawing premature conclusions.

{\it Acknowledgments} We thank Johanna Borissova, Fay Dowker, Stefano Liberati, Tony Padilla and Eleonora di Valentino for helpful discussions. 
RI was supported by a Bell-Burnell Fellowship and JM partly supported by STFC Consolidated Grant ST/T000791/1.

\clearpage

\appendix

\section{More on the Otto cycle and Fig.1}\label{Apcycles}
\begin{figure}[h!]
  \centering
  \includegraphics[width=0.5\textwidth]{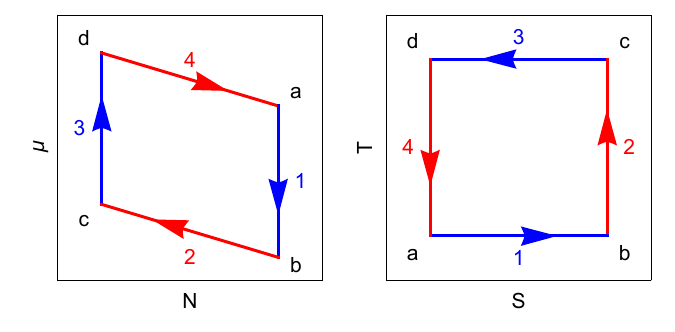}
  \caption{Sketch of an infinitesimal Otto cycle in the $(N,\mu)$ and $(S,T)$ planes generating $W>0$ but assuming $\tfrac{\partial T}{\partial N}=\tfrac{\partial \mu}{\partial S}<0$, in contrast with Fig.~\ref{FigOtto}. To obtain the corresponding cycles with $W<0$ one should reverse all arrows and swap labels 1 and 3 in this figure and in Fig.~\ref{FigOtto}. 
  }
  \label{FigOtto1}
\end{figure}

Unfortunately, the late-time acceleration requirement \(W>0\) (and hence \(T_{3}>T_{1}\)) does not fix the sign of \(\partial T/\partial N = \partial\mu/\partial S\). For definiteness, in Fig.~\ref{FigOtto}  and in the main text we adopted the positive sign, but \(W>0\) is equally possible (and the calculation proceeds with only obvious modifications) for \(\partial T/\partial N < 0\), with a cycle of the form shown in Fig.~\ref{FigOtto1}. In both Figs.~\ref{FigOtto} and \ref{FigOtto1}, reversing the arrows and exchanging the labels \(1\) and \(3\) takes us to the \(W<0\) cycle. We also assumed \(\partial\mu/\partial N < 0\) and \(\partial T/\partial S = 0\) in both Figures, although neither assumption is required for any of the arguments in this Letter.

In more general theories, legs 1 and 3 in the \((T,S)\) panel would be tilted (just as legs 2 and 4 are in the \((\mu,N)\) panel), but Eq.~\eqref{Q} still holds. This follows directly by transposing to this setting the derivation leading to Eq.~\eqref{Work}.

\section{Calculation of some integrals}

One can derive \eqref{lcavdef} by computing (writing $\hat{\mathbf{k}}$ in polar coordinates): 
\begin{equation}
    \int \frac{d^2\hat k}{4\pi}=1,\quad  \int \frac{d^2\hat k}{4\pi}k^i=0,\quad  \int \frac{d^2\hat k}{4\pi}k^ik^j=\frac{\delta^{ij}}{3}
\end{equation}
so that $\langle X_{\mu\nu} \rangle_{l.c.}=X_{00}+(X_{11}+X_{22}+X_{33})/3$. Noting that 
$\langle X_{\mu\nu} \rangle_{l.c.}$ is a scalar, but depends on how the directions $\hat{\mathbf{k}}$ are fixed by the mirror, and since the mirror's normal in this frame is $n^\mu=(1,\mathbf{0})$, we can write this expression as   \eqref{lcavdef}. 

The following integrals over the diamond or its boundary appear in the main text. Using \eqref{CKVF}, we have:
\begin{align}
        W&=-\int_{\mathcal{D}} dV \zeta_\nu\nabla_\mu T^{\mu\nu}
    \approx - \nabla_\mu T^{\mu0 }4\pi \int_0^\ell dr\int ^{\ell -r}_{-\ell+r }dt \, \chi_0\nn\\
    &=-\frac{8\pi }{45} \kappa \ell^5  n_\mu \nabla_\nu T^{\mu\nu} +\mathcal{O}(\ell^6)
\end{align}
with the other components of $\nabla_\mu T^{\mu\nu}$ dropping out because they multiply integrals with integrands odd in $t$, and the domain of the $t$-integration is symmetric. For the same reason ($\psi = -\kappa t/\ell$ is odd), the $\ell^{4}$ contribution to the conformal correction vanishes, leaving:
\begin{equation}
     \int_{\mathcal{D}} dV \psi T^\mu_\mu\approx -\dot T^\mu_\mu \frac{2\pi }{45}\kappa\ell^5+\mathcal{O}(\ell^6),
\end{equation}
(which is only needed to note the absence of the $\ell^{4}$ term).  

Finally, the leading order heat fluxes through $\mathscr{I}^\pm$ are the same and equal to:
\begin{align}
    \delta Q&=\int_{\mathscr{I}^-} d\Sigma_\mu \zeta _\nu T^{\mu\nu} \nn\\
    &=4\pi \kappa  \langle k^\mu k^\nu T_{\mu\nu }\rangle_{l.c.}\times \nn\\
    &\int _{-2\ell}^{0}d\lambda(\ell+\lambda/2)^2\lambda(1+\lambda/2\ell)+\mathcal{O}(\ell^5)\nn\\
    &=\frac{4\pi}{5} \kappa \ell^4  \langle k^\mu k^\nu T_{\mu\nu }\rangle_{l.c.}+\mathcal{O}(\ell^5)
\end{align}
where for definiteness we considered $\mathscr{I}^-$ and used $ d\Sigma_\mu = -dA d\lambda \, k_\mu$ and $\zeta^\mu = - \lambda (1+\lambda /2\ell)\kappa k^\mu$. 
Notice that, to leading order, computing the fluxes with $\chi^\mu$ instead of $\zeta^\mu$ would only change the result by a factor of 5/3.

Note also that for an infinitesimal Otto cycle, to leading order we can ignore variations in $\kappa$ in steps 2 and 4, which would lead to cubic order corrections. This is in keeping with ignoring variations of $\mu$ in steps 2 and 4 (see calculation leading to \eqref{Work}), or of $T$ in steps 1 and 3 if we considered theories with a dependence $S=S(T)$.

\section{On the Stueckelberg reinstatement}

As stated in the main text, 
``\dots as with any symmetry-breaking framework, one may restore invariance by introducing Stückelberg fields, but in extreme cases like ours this is not especially illuminating.''  
Indeed, even Galilean theory can be rendered diffeomorphism invariant by adding a sufficient number of such fields~\cite{Fried}.  This versatility of the procedure is its undoing, making it {\it by itself} physically vacuous. 

In our case, as in any framework partially anchored in Einstein gravity, one may always reconstruct the ``clipped'' equations in a reduced system (for example the trace subtracted in~\eqref{Eqenstracefree}) using the Bianchi identities.  
Specifically, 
\begin{align}
    G_{\mu\nu}-\tfrac{1}{4}g_{\mu\nu}G&=8\pi G_N \bigl(T_{\mu\nu}-\tfrac{1}{4}g_{\mu\nu}T\bigr),\\
       n_\mu \nabla_\nu T^{\mu\nu}&=
      -\frac{1}{L_2}\left(T_{\mu\nu}n^\mu n^\nu +\tfrac{1}{3}h^{\mu\nu}T_{\mu\nu}\right)
\end{align}
could be reconstructed as:
\begin{align}
     G_{\mu\nu}+\Lambda g_{\mu\nu}&=8\pi G_N T_{\mu\nu},\\
       n^\mu\partial_\mu \Lambda&=
    -\frac{8\pi G_N}{L_2}\left(T_{\mu\nu}n^\mu n^\nu +\tfrac{1}{3}h^{\mu\nu}T_{\mu\nu}\right)
\end{align}
exhibiting superficial similarities with interacting dark-energy models~\cite{IntDE,Eleanora}.  A preferred frame remains and could be removed by Stückelberg fields for \(n^\mu\), but beyond zeroth order the Stückelberg formulation becomes unwieldy, whereas the unsymmetric theory is simpler and clearer.

Even in an FRW setting, this is awkward. It gives
\begin{equation}
    \Lambda=\frac{4}{3}\!\left(\frac{1}{4L_2^{\,2}}+\frac{1}{L_2 t}\right),
\end{equation}
which diverges at the Big Bang, albeit more slowly than the usual matter components (which scale as \(1/t^2\)), so the standard early-time cosmology is not spoiled. 
More troubling is that acceleration requires matter creation and so a depletion of \(\Lambda\). If Lambda were to start from zero, it would be driven to negative values, but it starts from positive infinity. We regard these issues as shortcomings of the Stückelberg construction rather than of the proposed theory.

\section{Beyond the minimal theory; the example of anisotropy}
As an illustration of what happens beyond the homogeneity and isotropy of a FRW model we take the Bianchi I model: 
\begin{equation}\label{bianchiIapp}
   ds^{2} = -dt^{2} + A(t)^{2}dx^{2} + B(t)^{2}dy^{2} + C(t)^{2}dz^{2}. 
\end{equation}
This results in the four Einstein equations:
\begin{align}
\frac{\ddot{A}}{A} + \frac{\ddot{B}}{B} + \frac{\dot{A}\dot{B}}{AB} &= - 8\pi G_N p_1 , \\
\frac{\ddot{A}}{A} + \frac{\ddot{C}}{C} + \frac{\dot{A}\dot{C}}{AC} &= - 8\pi G_N p_2 , \\
\frac{\ddot{B}}{B} + \frac{\ddot{C}}{C} + \frac{\dot{B}\dot{C}}{BC} &= - 8\pi G_N p_3 , \\
\frac{\dot{A}\dot{B}}{AB} + \frac{\dot{B}\dot{C}}{BC} + \frac{\dot{A}\dot{C}}{AC} &= 8\pi G_N \rho,
\end{align}
where $p_i$ are the  diagonalized pressures. 
The trace-free Einstein equations 
reduce to three independent equations
capable of accommodating:
\begin{widetext}
\begin{equation}
    \dot \rho +\frac{\dot A}{A}(\rho+p_1)+\frac{\dot B}{B}(\rho+p_2)+\frac{\dot C}{C}(\rho+p_3)=\frac{1}{L_2}(\rho+\tfrac{1}{3}(p_1+p_2+p_3)).\nn
\end{equation}
\end{widetext}
By light cone averaging they further reduce to a single equation:
\begin{widetext}
    \begin{equation}
-\left(\frac{\ddot A}{A} + \frac{\ddot B}{B} + \frac{\ddot C}{C}\right)
+ \left(\frac{\dot A \dot B}{AB} + \frac{\dot B \dot C}{BC} + \frac{\dot A \dot C}{AC}\right)
= 4\pi G_{N} (3\rho + \sum_i p_i)\nn
\end{equation}
\end{widetext}
which would require two more non-conservation-like equations derived from the thermodynamics.

Even for vacuum, the solutions are different. For example it is possible to have solutions with $a_i \propto t^{\beta_i}$ which do not satisfy $\beta_1+\beta_2+\beta_3=1$ and $\beta_1^2+\beta_2^2+\beta_3^2=1$, so they do not need to be Kasner solutions. Indeed expansion in all directions is possible. An example is $\beta_1=\beta_2=1/2$ and $\beta_3=1+\sqrt{7}/2$.  


\begin{thebibliography}{99}
\bibitem{FN+}
Therefore, if when we change the temperature at entry and exit of a causal diamond, we are {\it not} claiming that the surface gravity on $\mathscr{I}^+$ and on $\mathscr{I}^-$ are different at the same external time.
\bibitem{FN1}
Our argument should not be confused with that in \cite{J2}, which focuses on the spatial $\Sigma$. We adopt the view that all entropy resides on the null surfaces.
\bibitem{FN2}The derivation in this paragraph can be redone with  the conformal KV, writing $\delta U=\delta Q = T_{\mu\nu}\zeta^\mu d\Sigma^\nu$ and evaluating geodesic deviation from $\delta \theta=\theta -\theta_0$, where $\theta_0=1/(\mp\ell +\lambda/2)$ (resulting in $A_0= 4\pi (\ell \mp \lambda/2)^2$) is the geodesic deviation before matter flows across $\mathscr{I}^\pm$ (describing a zeroth order entropy, to be subtracted).  Dropping the isochoric assumption on $\mathscr{I}^\pm$ will be examined in~\cite{thermoconsts}. Note that even within the isochoric assumption one could make other choices for the relation between curvature and entropy, e.g. based on the Wald entropy functional, leading to theories considered ``GR-like'' in this Letter; see \cite{Paddy1}.
\bibitem{FN8}
A species-dependent $L_2$ makes sense because
thermodynamics is generally non-linear, and $U$ is a non-linear function of $S$ just as  $R_{\mu\nu}$ is a non-linear function of $T_{\mu\nu}$.
\end{thebibliography}
\end{document}